\documentclass[conference]{IEEEtran}
\IEEEoverridecommandlockouts
\usepackage{cite}
\usepackage{amsmath,amssymb,amsfonts}
\usepackage{algorithmic}
\usepackage{graphicx}
\usepackage{textcomp}
\usepackage{xcolor}
\usepackage{caption}
\usepackage{subcaption}
\usepackage{hyperref}
\usepackage{cleveref}
\usepackage{booktabs}
	\usepackage{hyperref}
\hypersetup{
    colorlinks=true,
    linkcolor=blue,
    filecolor=magenta,      
    urlcolor=cyan,
    pdftitle={Overleaf Example},
    pdfpagemode=FullScreen,
    }

\usepackage{soul}

\def\BibTeX{{\rm B\kern-.05em{\sc i\kern-.025em b}\kern-.08em
    T\kern-.1667em\lower.7ex\hbox{E}\kern-.125emX}}
\begin{document}

\title{Tuning the feedback controller gains is a simple way to improve autonomous driving performance \\
\thanks{Ross Drummond was supported by a UK Intelligence Community Research Fellowship from the Royal Academy of Engineering.}
}

\author{\IEEEauthorblockN{Wenyu Liang,\, 
Pablo R. Baldivieso and Ross Drummond}
\IEEEauthorblockA{\textit{Dept. of Automatic Control and Systems Engineering} \\
\textit{University of Sheffield}\\
Sheffield, UK \\
\texttt{\{w.liang,\,p.baldivieso,\,ross.drummond\}@sheffield.ac.uk}}
\and
\IEEEauthorblockN{ Donghwan Shin}
\IEEEauthorblockA{\textit{Dept. of Computer Science} \\
\textit{University of Sheffield}\\
Sheffield, UK \\
\texttt{d.shin@sheffield.ac.uk}}
}

\maketitle

\begin{abstract}
Typical autonomous driving systems are a combination of machine learning algorithms (often involving neural networks) and classical feedback controllers.  Whilst significant progress has been made in recent years on the neural network side of these systems, only limited progress has been made on the feedback controller side. Often,  the feedback control gains are simply passed from paper to paper with little re-tuning taking place, even though the changes to the neural networks can alter the vehicle's closed loop dynamics. The aim of this paper is to highlight the limitations of this approach; it is shown that re-tuning the feedback controller can be a simple way to improve autonomous driving performance. To demonstrate this, the PID gains of the longitudinal controller in the TCP autonomous vehicle algorithm are tuned. This causes the driving score in \texttt{CARLA} to increase from  73.21 to 77.38, with the results averaged over 16 driving scenarios. Moreover, it was observed that the performance benefits were most apparent during challenging driving scenarios, such as during rain or night time, as the tuned controller led to a more assertive driving style. These results demonstrate the value of developing both the neural network and feedback control policies of autonomous driving systems \textit{simultaneously}, as this can be a simple and methodical way to improve autonomous driving system performance and robustness. 
\end{abstract}



\section{Introduction}

Training an Autonomous Driving System (ADS) requires solving a complex control problem so the autonomous vehicle can efficiently navigate through uncertain and changing environments. As a control problem, the development of an ADS is complicated by issues such as: \emph{i)} the  sensors (such as image data from cameras) are  only able to infer local and incomplete state information, \emph{ii)} the lack of clear definitions  for the control objectives and constraints to capture the diverse range of driving behaviours seen in practice,  and \emph{iii)} the ADS has to react to agents moving dynamically and, often,  aggressively. These issues represent a challenge that traditional control policies, such as model predictive control (MPC) \cite{mark}, struggle to handle. Instead, machine learning algorithms built from neural networks (NN) dominate the landscape of ADS. These data-driven NN algorithms are able to flexibly handle  both the wealth of data generated by the vehicle sensors as well as the vagueness of the control problem definition. They can now also be trained using readily-available  and simple software which do not require mathematical models, with this simplicity being another key reason fuelling the growth of NNs. Spurned on by the growth of NN control algorithms, ADSs have improved significantly in recent years, and there now exists an array of  open source software tools and challenges to evaluate different algorithms in benchmark scenarios. One notable example is the open source \texttt{CARLA} driving simulator \cite{carla} that hosts the  \texttt{CARLA} Autonomous Driving Leaderboard \cite{22}. \texttt{CARLA} is used in this paper to evaluate the ADS algorithms. 

Currently, the algorithms dominating the \texttt{CARLA} Leaderboard are those based on NNs that translate the vehicle's image/lidar data into control actions for steering, throttle and braking. Examples include: Trajectory-guided control prediction (TCP) \cite{tcp}, ReasonNet \cite{ReasonNet}, Interfuser \cite{interfuser}, and Pylot \cite{pylot}, with \texttt{CARLA} also being used as a resource for software testing ADSs \cite{haq2022efficient}. However, while such NN-dominated ADS control algorithms have performed well in experiments, their limitations affect how well they perform in the field. Most significant are the issues around robustness and explainability, which are, coincidentally, two issues which classical control algorithms can, arguably, deal with, at least in principle. However, the rate of progress of NN-based ADS algorithms has meant that, often, there is not enough time/data to follow the classical control design pipeline of developing a mathematical model, tuning the controller, and then verifying its robustness \emph{e.g.} using Lyapunov functions. Instead, different NN architectures are proposed and trained using the latest techniques where only limited stability analysis/modelling/testing is conducted. As such, to keep pace with the latest NN developments, often the control gains used to manipulate the vehicle towards the reference trajectory generated by the NN are simply passed from paper to paper without re-tuning. An example of this is the Proportional-Integral-Derivative (PID) control gains of \cite{tcp} which were carried over from \cite{pid_orig}, as the tuning from  \cite{pid_orig} was deemed satisfactory. From a control perspective, this practice of carrying over PID gains is problematic. Changes to the NN in the ADS can impact the vehicle's closed-loop dynamics, and this change should be accounted for in the controller gains. 

The aim of this paper is to demonstrate the value of tuning the feedback controller gains in ADS which use NNs to generate the vehicle's reference trajectories. Whilst rapid advances have been for the NNs of ADS, often little attention is paid to the re-tuning of the controller gains. It is shown here that re-tuning these gains can be a simple way to improve ADS performance. And, since the tuning can be done using the methods and machinery of control theory, the resulting improvements can be implemented at scale and in a methodical way, potentially with robustness guarantees as well \cite{drummond2022bounding}. In this way, the results of this paper are aimed at opening a dialogue between the control theory and machine learning methods behind ADSs. Specifically, our work on re-tuning ADS controllers have led to the following results:
\begin{enumerate}
    \item Re-tuning the PID gains in the TCP \cite{tcp} autonomous driving systems algorithm can improve performance.
    \item The results are evaluated in \texttt{CARLA}, with the driving score of 73.21 for TCP increased to 77.38 by simply tuning the gains of its longitudinal PID controller.
    \item The performance improvements of the tuned controller were most apparent in low-light and wet scenarios.
\end{enumerate}
These results are preliminary and further improvements are expected through the application of more advanced tuning methods, such as automatic tuning  \cite{aastrom2006advanced} or  designing the controller to avoid overshooting  \cite{taghavian2021pole}. Given the promising results obtained, we expect that employing more advanced control methods for ADS will lead to further improvements in the driving score. In particular, there is great promise in the application of MPC  for this problem, to take advantage of the future way points generated by NNs planning algorithms such as TCP. Combining those future way points with the robustness and constraint satisfaction feature of MPC should deliver further improvements in the driving score.

In terms of comparison to the literature, there is a wealth of results and methods in the area of ADS, as reviewed in papers such as \cite{22,36} In this work, the focus is on improving the driving score of the TCP ADS described in \cite{tcp}  for \texttt{CARLA} \cite{carla}. The essence of our result lies in showing how tuning the feedback controller gains, here the focus is on the the longitudinal PID controller's gains, in an ADS which uses NNs to generate reference trajectories can  improve performance. As mentioned above, many of the emerging NN-based algorithms do not re-tune the feedback controller gains during development, an issue which could deteriorate performance and/or increase training time. Our aim is to highlight this limitation and recommend that a dialogue should be opened between machine learners and control engineers to remedy it, as doing so can bring quick and simple gains in the driving score. We focus on the TCP algorithm for ADS in this work, but we are conscious that the idea of the need for controller re-tuning should be applicable to a broad class of ADS algorithms. Moreover, algorithms for vehicle control have also been implemented from the pure control theory perspective, for example the eco-driving results of \cite{Moura},  the robust MPC approach developed in \cite{paul} or the stochastic MPC approach of \cite{nair2023predictive}. The difference with these more control-focused results, such as \cite{paul}, is that whilst they are often equipped with performance and optimality guarantees, a feature lacking for many NN algorithms, e.g. \cite{pylot} and \cite{tcp}, these algorithms often require full state/road information which may not always be available

\section{CARLA Driving Simulator}

The results of this paper are generated with the \texttt{CARLA} driving simulator, with version \texttt{CARLA 0.9.10} run on Linux. \texttt{CARLA} is as an open-source driving simulator based upon \texttt{Unreal Engine 4}  able to create a diverse range of driving scenarios for training and validating ADS. These scenarios include different weather settings, city infrastructure, traffic density, and other such dangerous driving scenarios that are impractical to replicate within the real physical world.  Environments in \texttt{CARLA} consist of 3D representations of both static, \emph{i.e.,} buildings, vegetation, infrastructure, or dynamic elements, \emph{i.e.,} moving traffic or pedestrians. Additionally, \texttt{CARLA} offers an array of sensors for collecting data, such as RGB cameras, LiDAR, radar, GPS, IMU, speedometers, and pseudo-sensors that provide semantic segmentation and ground-truth depth information \cite{carla}. 

\texttt{CARLA} also hosts the Autonomous Driving Challenge Leaderboard to assess the driving capabilities of autonomous agents within real traffic scenarios. Throughout a series of predetermined routes, the ego vehicle encounters a diverse array of traffic situations, including lane merging, lane changing, interactions at crossroads and roundabouts, responses to traffic lights and signs, and interactions with bicycles and pedestrians.

\subsection{Driving score}
Given an autonomous vehicle, the performance of its ADS can be evaluated in \texttt{CARLA} using the Driving Score of the \texttt{CARLA} Driving Challenge. To generate the score, the ADS undergoes evaluation across 16 route scenarios. Each route scenario is repeated three times, and the results for these three routes are then averaged. When an infraction occurs during a run, the penalty coefficients of Table \ref{table:infractions} are applied. At the culmination of the route, regardless of whether the ego vehicle has reached its destination or not, the total penalty score is determined by multiplying all the penalty coefficients together. A penalty score of 1 is registered if no infractions occurs. 

\begin{table}[h!]
\centering
\caption{Penalty coefficient for the infractions.}
\label{table:infractions}
\begin{tabular}{ll} 
\toprule
  Infraction & Coefficient\\
\midrule
 Collision with pedestrian & 0.5 \\
  Collision with pedestrian & 0.6 \\
   Collision with pedestrian & 0.65 \\
    Running a red light & 0.7 \\
Running a stop sign & 0.8 \\
Off-road driving & 1-Percentage of the completed route \\
\bottomrule
\end{tabular}
\end{table}

Furthermore, if the agent's progress is halted, then the simulation has to be suspended as detailed in Table \ref{table:shutdown}. In such instances, the current score at the time of the halting is recorded and the evaluation proceeds to the next scenario. 

\begin{table}[h!]
\caption{Shut down events}
\label{table:shutdown}
\begin{center}
\begin{tabular}{ll}
\toprule
 Shut down event & Explanation\\
\midrule
 Route deviation & More than 30m deviation from the planned route \\
  Agent blocked & Agent no action for 180s \\
    Simulation timeout & No client-server communication after 60s \\
Route timeout & Simulation route takes too long to finish \\
\bottomrule
\end{tabular}
\end{center}
\end{table}

The ADS's \textbf{Driving score} is the route completion score multiplied by the total penalty score, where
\begin{enumerate}
    \item \textbf{Total penalty score}: Obtained by multiplying all the penalty coefficients together.
    \item \textbf{Route completion} is the percentage of the route's length that an agent has completed. A successful arrival at the destination corresponds to a perfect score of 100.
\end{enumerate}

After completing all the routes, an overall infraction score is derived by normalizing against the infraction count incurred per kilometre travelled. This is achieved by subtracting the average of all penalties from 1. Similarly, the global route completion score is the mean for all routes, and the global driving score  is the global infraction score multiplied by the global route completion score.


\section{Autonomous Driving Systems: TCP}

\begin{figure}
         \centering
         \includegraphics[width=0.48\textwidth]{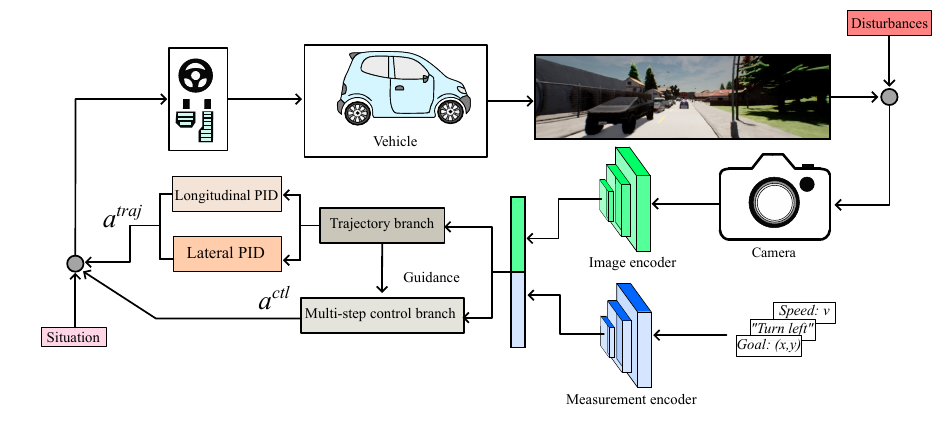}
         \caption{Schematic of the TCP algorithm from \cite{tcp}.}
         \label{fig:tcp_illustration}
\end{figure}

In this work, the TCP algorithm from \cite{tcp} is used as the benchmark ADS. The structure of TCP is shown in Figure \ref{fig:tcp_illustration} and it splits into two different branches,  a \textbf{Trajectory branch} and a \textbf{multi-step control prediction branch}. This combination contrasts with most end-to-end driving systems, e.g. \cite{bojarski2016end}, which typically either exclusively map raw sensor data to   trajectories  \textit{or} low-level control actions. 

\subsection{Trajectory planning branch}
 Within TCP's trajectory planning branch, images captured by the camera are processed by a CNN-based image encoder to generate a feature map. Simultaneously, measurement data, comprising information such as location, speed, and navigation is processed by a multi-layer NN measurement encoder to generate measurement features. These features are then amalgamated via average pooling, forming a trajectory feature that enters a Gated Recurrent Unit (GRU) \cite{32} to predict four waypoints for the vehicle to follow.

 These waypoints form a reference trajectory for the autonomous vehicle. PID controllers for both the longitudinal and lateral direction are then used to regulate the autonomous vehicle and follow this reference. The longitudinal controller is used to generate throttle and brake control actions whilst the lateral controller is used to implement steering action. In this work, the focus will be on tuning the longitudinal controller, with the gains of the lateral controller kept the same as in \cite{tcp}. The values of the longitudinal PID gains are stated in Table \ref{table:gains} 
with $k_p$ being the proportional gain, $k_d$ the derivative gain  and $k_i$ the integral gain.

\begin{table}
\caption{Values for the longitudinal PID control gains of both the original and tuned versions of TCP \cite{tcp}. Here, $k_p$ is the proportional gain, $k_i$ is the integral gain and $k_d$ is the derivative gain. }
\begin{center}
\begin{tabular}{c|c|c|c|c|c} 
   & $k_p$ & $k_i$ & $k_d$ & Max throttle & Brake speed \\
  \hline
 TCP & 5 & 0.5 & 1 & 0.75 & 0.4 \\
 TCP-tuned & 11 & 0.1 & 1 & 0.8 & 0.45 
\end{tabular}
\end{center}
\label{table:gains}
\end{table}

\begin{figure*}
     \centering
     \begin{subfigure}[b]{0.49\textwidth}
         \centering
         \includegraphics[width=\textwidth]{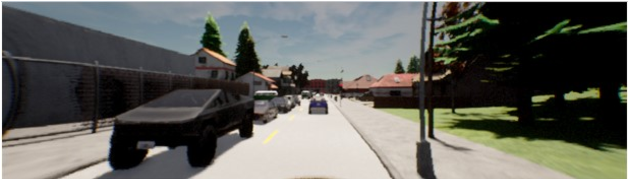}
         \caption{\textit{Clear Noon} in \textit{Town02}.}
         \label{fig:y equals x}
     \end{subfigure}
     \hfill
     \begin{subfigure}[b]{0.49\textwidth}
         \centering
         \includegraphics[width=\textwidth]{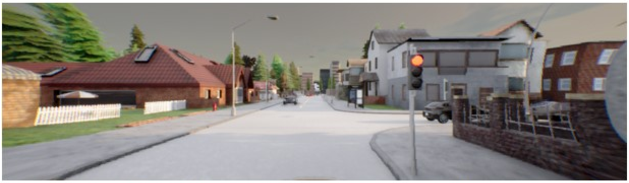}
         \caption{\textit{Cloudy Sunset} in \textit{Town02}.}
         \label{fig:three sin x}
     \end{subfigure}
     \hfill
     \begin{subfigure}[b]{0.49\textwidth}
         \centering
         \includegraphics[width=\textwidth]{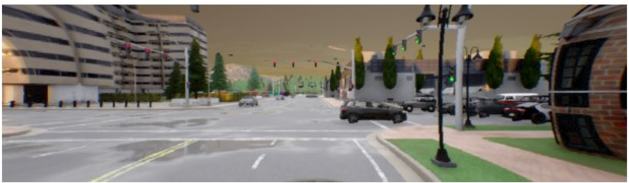}
         \caption{\textit{Soft Rain Dawn} in \textit{Town05}.}
         \label{fig:five over x}
     \end{subfigure}
          \begin{subfigure}[b]{0.49\textwidth}
         \centering
         \includegraphics[width=\textwidth]{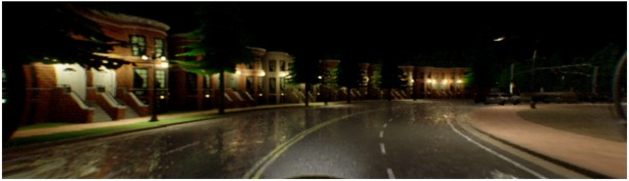}
         \caption{\textit{Hard Rain Night} in \textit{Town05}.}
         \label{fig:five over x}
     \end{subfigure}
        \caption{Images from the drivers perspective of the vehicle in four different weather scenarios.}
        \label{fig:weather}
\end{figure*}

\begin{figure}
         \centering
         \includegraphics[width=0.4\textwidth]{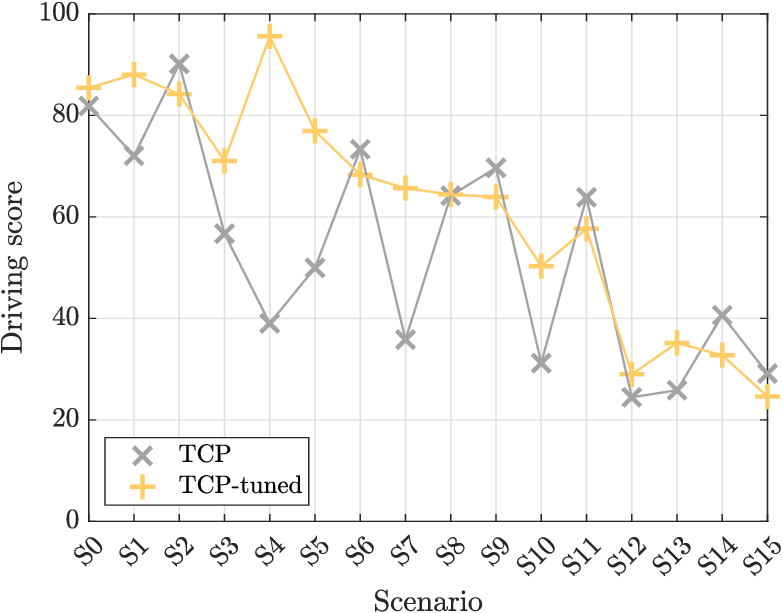}
         \caption{Comparison between the original driving scores for the original TCP algorithm~\cite{tcp} and that with the tuned PID gains for the longitudinal controller, referred to as TCP-tuned. }
         \label{fig:scores}
\end{figure}

\subsection{Multi-step control  branch }
In the multi-step control prediction branch of TCP, the goal is to directly forecast control actions for the ensuing four time steps. Within this branch, an embedded temporal module is used to implicitly account for the dynamic interactions and alterations within the environment and the agent. This module aims to predict how objects move within the environment and relative to the ego vehicle, and so anticipate dynamic changes.

\subsection{Integrating the  branches }

The control actions from both the trajectory branch and the multi-step control branch  of TCP \cite{tcp} are then combined together using a situation-based fusion strategy. Specifically, a situation is determined by calculating the steering actions over the previous seconds. If more than half of these steering actions surpass 0.1, then the vehicle is assumed to be turning and a ``\textit{control specialized}" situation is regarded as being active. Conversely, if this criterion is not met, then the situation is deemed ``\textit{trajectory specialized}." A weight, $\alpha \in [0, 0.5]$, is used to combine the two  control actions from the trajectory branch, $a^\text{traj}[k]$, and the multi-step control prediction branch, $a^\text{ctl}[k]$, to generate the final control action. When the vehicle is in a control specialised state, the combined action, $a[k]$, is
\begin{align*}
    a[k] = \alpha \times a^\text{ctl}[k] + (1-\alpha)\times a^\text{traj}[k]
\end{align*}
and, when it is trajectory specialised, then the two control actions are combined according to
\begin{align*}
    a[k] = \alpha \times a^\text{traj}[k] + (1-\alpha)\times a^\text{ctl}[k].
\end{align*}
In this paper, the weight was set to $\alpha = 1/2$.

\subsection{Tuning the Feedback control algorithm}

The values of the longitudinal PID gains from TCP \cite{tcp} and the tuned values considered in this paper are shown in Table \ref{table:gains}---the tuned gains are referred to as TCP-tuned. No other additions were made to TCP beside these changes to the longitudinal PID gains. Simple trial-and-error tuning was applied to tune the gains, with it being observed that a reduction in the integral gain $k_i$ and an increase in the proportional gain $k_p$ led to a more reactive and dynamic driving resposne. It is acknowledged that this is rudimentary tuning compared to, for instance, the methods detailed in \cite{aastrom2006advanced}; further improvements over the results of Table \ref{table:score} are expected using more effective tuning algorithms, and even more advanced control algorithms such as MPC. However, the results of this paper demonstrate that even rudimentary tuning of the PID gains can improve the results, as will be seen in the following section.

\section{Results}

\begin{figure*}
     \centering
     \begin{subfigure}[b]{0.49\textwidth}
         \centering
         \includegraphics[width=\textwidth]{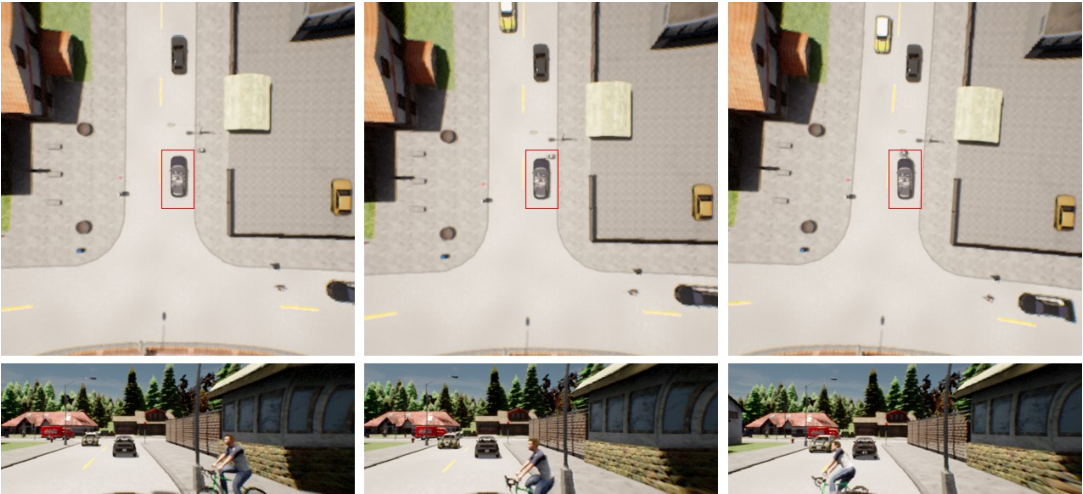}
         \caption{Collision with a bicycle.}
         \label{fig:bike}
     \end{subfigure}
     \hfill
     \begin{subfigure}[b]{0.49\textwidth}
         \centering
         \includegraphics[width=\textwidth]{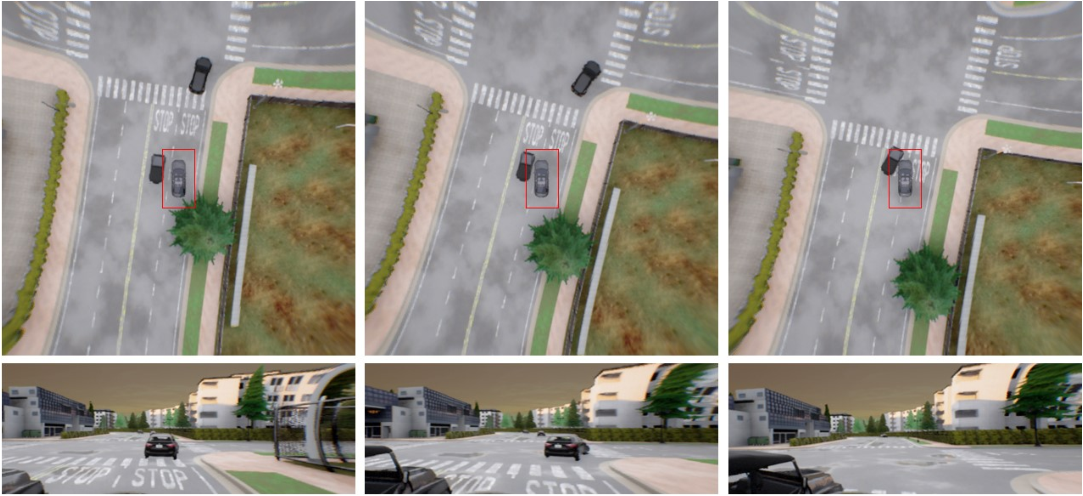}
         \caption{Collision with a vehicle when turning.}
         \label{fig:turning_tcp}
     \end{subfigure}
     \hfill
     \begin{subfigure}[b]{0.49\textwidth}
         \centering
         \includegraphics[width=\textwidth]{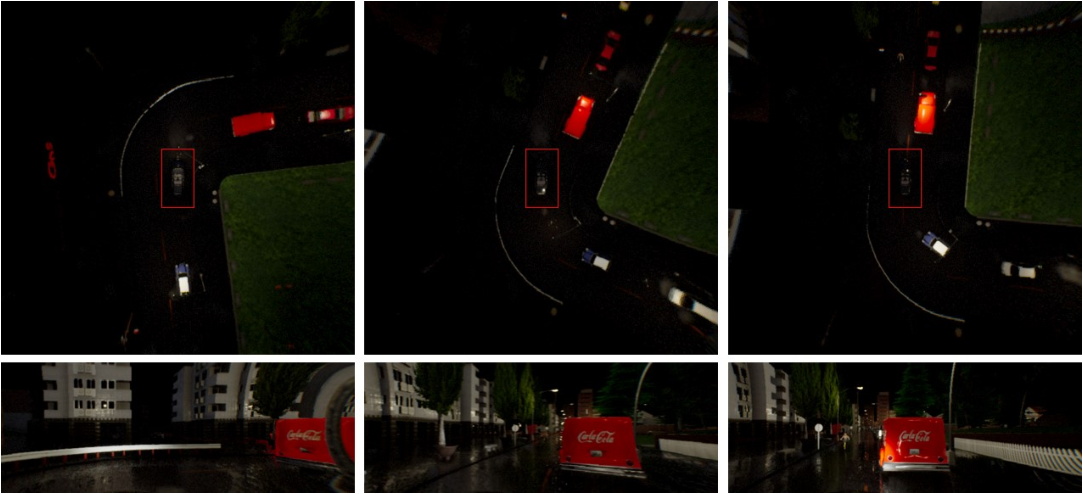}
         \caption{Deviating from the lane.}
         \label{fig:five over x}
     \end{subfigure}
          \begin{subfigure}[b]{0.49\textwidth}
         \centering
         \includegraphics[width=\textwidth]{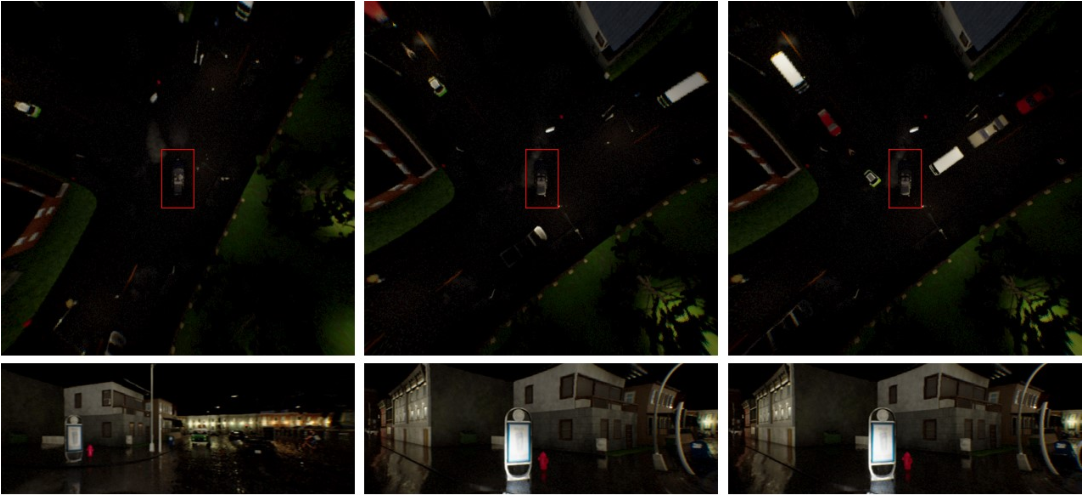}
         \caption{Blocked by an advertising board.}
         \label{fig:five over x}
     \end{subfigure}
        \caption{Four examples of crashes recorded with the TCP \cite{tcp} algorithm in CARLA (ego vehicle highlighted by red  box). }
        \label{fig:crashes}
\end{figure*}

\begin{figure*}
     \centering  
     \begin{subfigure}[b]{0.49\textwidth}
         \centering
         \includegraphics[width=\textwidth]{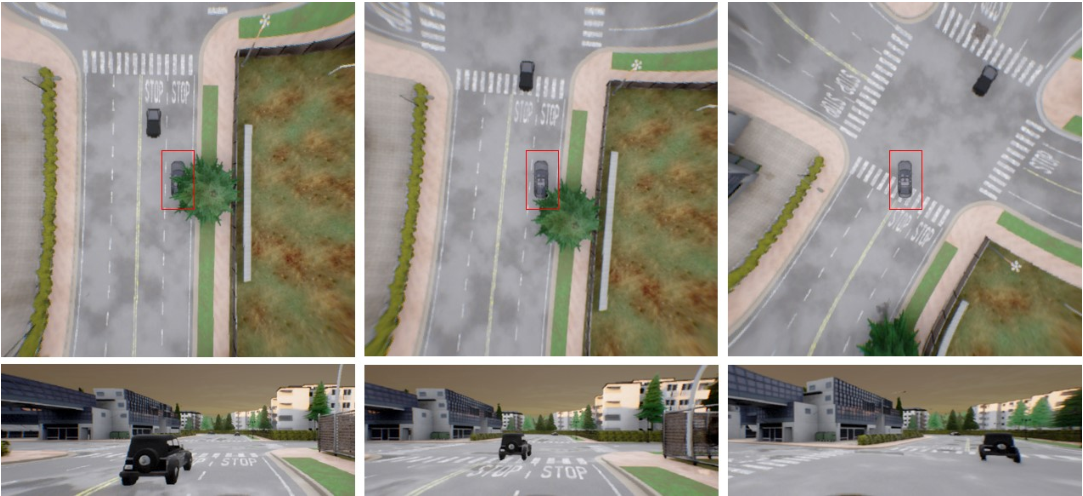}
         \caption{Collision avoided with vehicle at intersection.}
         \label{fig:turning_nope}
     \end{subfigure}
     \hfill
      \begin{subfigure}[b]{0.49\textwidth}
         \centering
         \includegraphics[width=\textwidth]{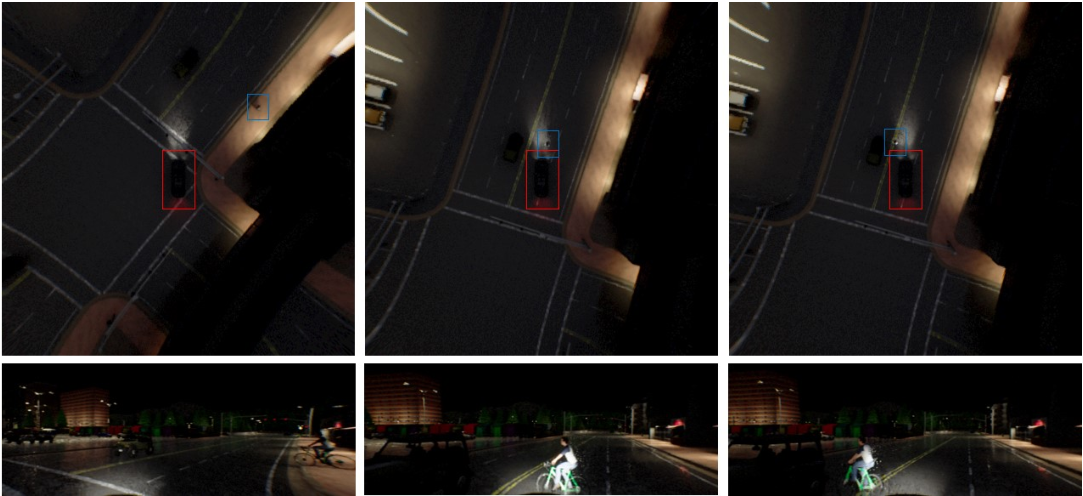}
         \caption{Collision avoided with bicycle. }
         \label{fig:bike_nope}
     \end{subfigure}
        \caption{Instances of the TCP algorithm with tuned PID gains avoiding crash events. In both cases, the ego vehicle is highlighted by a red bounding box. The bicycle in b) is highlighted by a blue bounding box. }
\end{figure*}

This section compares the \texttt{CARLA} driving score results of the  original TCP algorithm of \cite{tcp} against the version with the tuned PID gains.  Again, this tuned TCP algorithm is referred to as TCP-tuned in the results. The score is based upon the analysis of 4 different routes, 2 from \textit{Town02} and 2 from \textit{Town05}. \textit{Town02} features T-junctions in a small town setting, while \textit{Town05} represents a larger city with multiple traffic lanes. Both towns are provided in CARLA by default. Each route is assessed under four distinct weather conditions (\textit{Clear Noon}, \textit{Cloudy Sunset}, \textit{Soft Rain Dawn}, \textit{Hard Rain Night}). Figure \ref{fig:weather} provides a visual representation of different weather conditions and town scenarios as viewed by the driver.

The combination of the four different routes with the four different weather events resulted in a total of 16 different route scenarios being tested. These sixteen scenarios are labelled $S0-S15$ and the results comparing the original TCP algorithm with the tuned version is shown in Figure \ref{fig:scores}.  In each scenario, the traffic participants were randomly generated.

\begin{table}
\caption{Penalty coefficient of the infractions for both TCP and the version with the tuned PID gains used in this work.}
\begin{center}
\begin{tabular}{cccc} 
\toprule
   & Driving score & Route completion & Infraction penalty \\
\midrule
 TCP & 73.21 & 85.63 & 0.855 \\
 TCP-tuned & 77.38 & 89.36 & 0.866 \\
\bottomrule
\end{tabular}
\end{center}
\label{table:score}
\end{table}

\subsection{Driving Score}
Table \ref{table:score} compares the driving score for the original TCP and that with the tuned longitudinal PID gains from Table \ref{table:gains}. The results show that the tuned version of TCP can exhibit superior performance when evaluated across all sixteen driving scenarios, with the PID tuning increasing the Driving Score from 73.21 to 77.38. It was noted that that TCP suffered particularly in the challenging driving scenarios of  \textit{Hard Rain Night}, with the reduced visibility and the ground reflections impacting the camera sensors and, correspondingly, TCP's ability to maintain a precise lane position for the vehicle.  It was observed that the more assertive driving of the tuned TCP algorithm (with a higher proportional PID gain and a lower integral one) was able to alleviate some of the hesitancy of the driver in these challenging driving scenarios, and hence improve the score. To reduce the risk of this assertive driving style leading to more crashes,  the brake ratio of TCP \cite{tcp} was increased to force the  vehicle to initiate braking earlier.

The line chart of Figure \ref{fig:scores} compares the driving scores for both the original version of TCP \cite{tcp} against the one  with the tuned PID gains. The figure demonstrates that the tuned controller can overcome some of the challenging scenarios, such as avoiding the collision with the bicycle in Scenario 4 and passing the blocking point in Scenario 7, resulting in a higher driving score in these scenarios.

\subsection{Inspection of crash incidents}

Tuning the feedback PID gains in the TCP not only increased the overall driving score but also meant that certain crash situations could be avoided. 
Figures \ref{fig:crashes} shows several scenarios of the autonomous vehicle crashing in \texttt{CARLA} when controlled by the untuned TCP algorithm. These examples include the vehicle colliding with a bicycle which unexpectedly crossed the road, crashing into a vehicle when turning around a corner, leaving its lane when turning and stopping for some unknown reason when encountering an advertisement board.  In particular, it was observed that the vehicle stopped at the advertisement board on all three repetitions of the route, which was on a rainy night. Videos showing simulations with TCP can be found 
in \cite{Drummond2023_1}
Specifically, the files \texttt{Original$\_$S7$\_$1.mp4}, \texttt{Original$\_$S7$\_$2.mp4} and \texttt{Original$\_$S7$\_$3.mp4} show the TCP-driven vehicle being blocked twice at an advertising board and exceeding the allotted time once.
  
 %
Often these collision incidents could be avoided with the tuned version of TCP. For instance, the collision observed in Scenario 10 of the vehicle crashing when turning round a corner, as shown in Figure \ref{fig:turning_tcp}, was  avoided after tuning the PID gains, as illustrated in Figure \ref{fig:turning_nope}. In that instance, the ego vehicle maintained some distance from the adjacent vehicle and the stop line, allowing it to navigate through without colliding. The ego vehicle with the tuned gains also demonstrated an ability to evade bicycles crossing the road unexpectedly, even in challenging conditions like heavy rain at night. An example of this is shown in Figure \ref{fig:bike_nope} where, despite the bicycle being barely discernible to human eyes, the ego vehicle halted in a timely manner, avoiding a collision. Videos of the vehicle driven by TCP with the tuned PID gains can be found 
in \cite{Drummond2023}. In this case, the collision in Scenario 10 from TCP  shown in Figure \ref{fig:turning_tcp} is avoided, as shown in Figure \ref{fig:turning_nope}. The video of the vehicle evading the bicycle, as shown in Figure \ref{fig:bike_nope}, can be found in the file \texttt{TCP3$\_$S10$\_$2.mp4}. 

\section*{Conclusions}
It was shown that simply re-tuning the feedback control gains embedded within neural-network based autonomous driving systems can improve performance, especially for challenging driving scenarios such as when it is raining and during night time. To demonstrate this result, the PID gains for the longitudinal control of the TCP algorithm were tuned and simulations were conducted in \texttt{CARLA} to evaluate performance. After tuning  the PID gains, the driving score over 16 scenarios  was increased from 73.21 to 77.38-- demonstrating that tuning the control gains can be a simple way to improve driving performance. The aim of this work was to highlight that while there has been rapid advancement in neural network design for autonomous driving systems, there is value in spending time and effort in tuning the feedback control gains as well. We hope that this result can encourage a dialogue between machine learners and control engineers to create more robust and explainable driving systems  which retain the flexibility and power of the underlying machine learning algorithms.

We are conscious that the presented results are rudimentary (manual tuning of a PID controller is studied) and  we see significant opportunity to develop these results further. Specifically, we see potential in applying more advanced tuning algorithms and control policies (in particular MPC since the future way points generated by TCP can be embedded within the MPC cost). Moreover, we believe that these results would benefit from the certificates of neural network robustness. However, whilst guaranteeing closed-loop stability of the system is an appealing objective from a control theory point of view, it was observed that most of the crashes encountered in the \texttt{CARLA} simulations could be attributed to poor trajectories being generated (i.e. pushing the vehicle towards a crash/leaving a lane) rather than some underlying instability of the closed-loop feedback system. Still, by applying the deep mathematical toolbox of  control theory, we expect significant benefits in terms of robustness and explainability could  be achieved.

\bibliographystyle{IEEEtranS}
\bibliography{bibliog.bib}

\end{document}